\documentclass{PoS}
\usepackage{epsfig}

\PoS{PoS(HEP2005)338}

\title{Discovery Potential of MSSM Higgs Bosons with ATLAS}

\ShortTitle{Discovery Potential of MSSM Higgs Bosons with ATLAS}

\author{\speaker{Johannes Haller}\thanks{on behalf of the ATLAS collaboration}\\

        CERN, Switzerland\\

        E-mail: \email{haller@mail.cern.ch}}

\abstract{In this article the potential of the ATLAS experiment to discover MSSM Higgs bosons is discussed. Various Monte-Carlo studies for SM Higgs boson production and dedicated MSSM Higgs boson analyses are taken into account to investigate the discovery potential in four different CP-conserving MSSM benchmark scenarios, the $M_h$--max scenario, the no mixing scenario, the gluophobic scenario and the small $\alpha$ scenario. The most recent theoretical calculations are used for the prediction of Higgs masses and couplings. The results are shown for integrated luminosities of 30 and 300 inverse fb. With 300 inverse fb, a five sigma discovery of at least one Higgs boson state is possible in all scenarios for all values of $\tan\beta$ and $M_A$. The heavy neutral and charged Higgs states can only be discovered in parts of the parameter space.}

\FullConference{International Europhysics Conference on High Energy Physics\\

        July 21st - 27th 2005\\

        Lisboa, Portugal}

\begin{document}

\section{Introduction}
Supersymmetric extensions (SUSY) of the Standard Model (SM) are considered to be the most promising theories describing new physics beyond the SM. SUSY particles are predicted with masses below $\sim 1 $\,TeV. Their direct production has been searched for at all high-energy particle colliders, most recently at LEP, HERA and the Tevatron, leading to no discovery so far. The LHC has excellent perspectives to discover SUSY due to its large centre-of-mass energy. In this article we present the potential of the ATLAS experiment at the LHC for the discovery of Higgs bosons in various scenarios of the minimal SUSY extension of the SM (MSSM).

\section{MSSM Higgs Sector and Benchmark Scenarios}
In the MSSM two complex Higgs field doublets are needed in order to give masses to the fermions without introducing anomalies. Assuming CP-conservation, this leads to five physical Higgs bosons: three neutral states -- the two CP even $h$ and $H$ and the CP odd $A$ -- and two charged states $H^{\pm}$.  At Born level the phenomenology of the Higgs sector is determined by just two parameters, usually chosen to be the ratio of the vacuum expectation values of the two doublets $\tan\beta$ and the Higgs mass $M_A$. At least one of the neutral Higgs bosons is predicted to have a mass below the $Z$ mass. However, radiative corrections shift this limit to about 135\,GeV~\cite{SVEN} for a top quark mass $M_{\rm top}= 175$\,GeV and a soft SUSY-breaking parameter in the sfermion sector at the electroweak scale $M_{\rm SUSY}=1$\,TeV. The corrections mainly stem from the $t/\tilde{t}$ sector and therefore the most important parameters are $M_{\rm top}$, $M_{\rm SUSY}$, the stop mixing parameter $X_t$, the SU(2) gaugino mass at the electroweak scale $M_2$, the SUSY Higgs mass parameter $\mu$ and the gluino mass $M_{\rm gluino}$.
Rather than varying all of the above parameters independently, usually benchmark scenarios are investigated where $\tan \beta$ and $M_A$ are scanned while the other parameters are fixed. The scenarios discussed in this article have been suggested~\cite{CARENA} in order to examplify the LHC discovery potential. The $M_h$-max benchmark set gives the largest value for $M_h$, corresponding to the most conservative exclusion from LEP~\cite{LEPHIGGS}, whereas the no mixing scenario leads to a small value for $M_h$. The gluophobic scenario has been designed to suppress the coupling of $h$ to gluons affecting the LHC discovery of the channels in gluon-gluon fusion (GGF). In the small $\alpha$ scenario the $h$ couplings to $b$ and $\tau$ are reduced mainly affecting the channel $h\rightarrow \tau\tau$ in Vector Boson Fusion (VBF) and the Higgs channel with associated top quark production $tth$ with $h\rightarrow bb$. 

\section{Monte-Carlo Studies}
The discovery potential reported here has been obtained from the latest results~\cite{TDR,MCSTUDIES} of Monte-Carlo (MC) studies for the SM Higgs boson and dedicated MSSM Higgs MC analyses\footnote{
The following channels have been considered: for neutral Higgs bosons ($\cal H$): VBF with ${\cal H}\rightarrow \tau \tau, WW$ and $\gamma\gamma$, $tt\cal{H}$ with ${\cal H} \rightarrow bb$, ${\cal H}\rightarrow \mu \mu$ and ${\cal H}\rightarrow \tau \tau$ from GGF and $bb{\cal H}$, ${\cal H} \rightarrow \gamma \gamma$ from GGF, $W\cal H$ and $tt\cal H$, ${\cal H}\rightarrow ZZ\rightarrow 4l$ and ${\cal H}\rightarrow WW \rightarrow l \nu l \nu$ from GGF, $W\cal H$ with ${\cal H}\rightarrow WW \rightarrow l\nu l\nu$, $H/A$ with $H/A\rightarrow tt$, $H\rightarrow hh\rightarrow \gamma\gamma bb$ and $A\rightarrow Zh\rightarrow llbb$. For charged Higgs bosons: $gb\rightarrow tH^{\pm}$ with $H\rightarrow tb$ and $H\rightarrow\tau\nu$ and in the decay of top quarks $pp\rightarrow tt\rightarrow bWbH^{\pm}$ with $H\rightarrow \tau\nu$.  
}. Compared to previous ATLAS MSSM Higgs discovery potential studies~\cite{TDR} the Higgs decay to $\tau\tau$, WW and $\gamma \gamma$ in VBF and the channel $t t \rightarrow  bW+bH \rightarrow bqq+b \tau \nu$, as well as double hadronic $\tau$ decays in $bbH\rightarrow H\rightarrow \tau\tau$ are newly included. Better background simulations are used for the channel $ttH$ with $H\rightarrow bb$. In all studies key performance numbers (e.g. trigger efficiencies,  b-tagging, $\tau$-identification, mass resolutions, etc.) have been obtained from a full simulation of the ATLAS detector. However, the signal selection efficiencies and the background expectations are derived from fast simulations. The systematic uncertainties of the MC studies are not included so far. For the calculation of the ATLAS discovery\footnote{A discovery means that the probability of a background fluctuation to the number of expected signal+background events is less than $2.85\times 10^{-7}$ using Poissonian statistics.} potential presented here masses and couplings are calculated using the most recent theoretical calculations as implemented in the FeynHiggs software package~\cite{FEYNHIGGS}. Leading order cross sections have been used for all production processes using a top quark mass of 175\,GeV. For the neutral Higgs boson production the SM like calculations from \cite{SPIRA} with appropriate MSSM corrections have been used. The production cross section of charged Higgs bosons via $bg\rightarrow tH^{\pm}$ has been calculated following~\cite{PLEHN}. PYTHIA~\cite{PYTHIA} was used for the process $tt\rightarrow bbWH^{\pm}$.

\section{Results on the Discovery Potential}
\begin{figure}[tb]
 	\begin{picture}(100,170)
	  \put(20,10){\epsfig{file=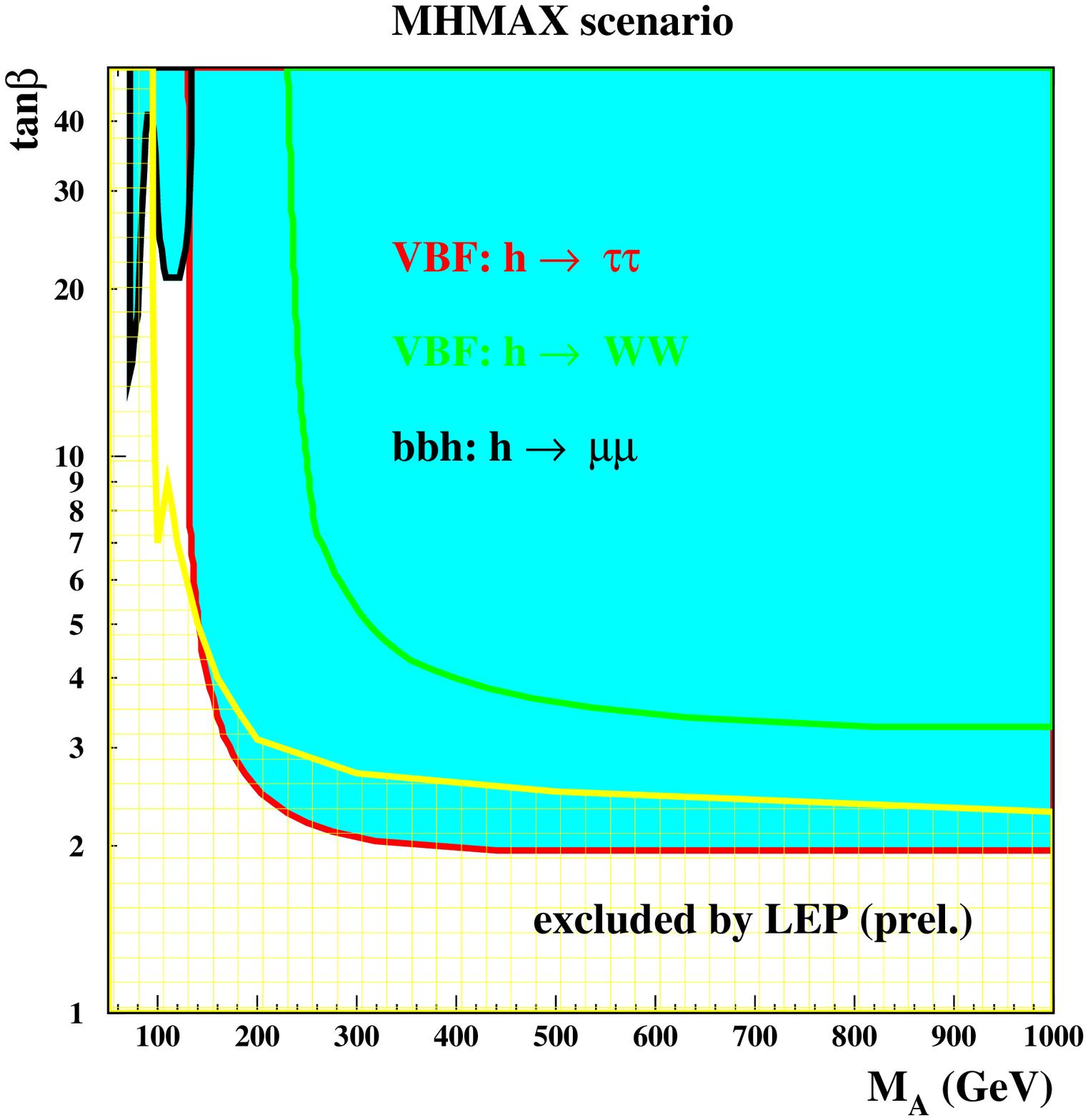,scale=.35}}
	  \put(105,90){ATLAS preliminary}
	  \put(220,10){\epsfig{file=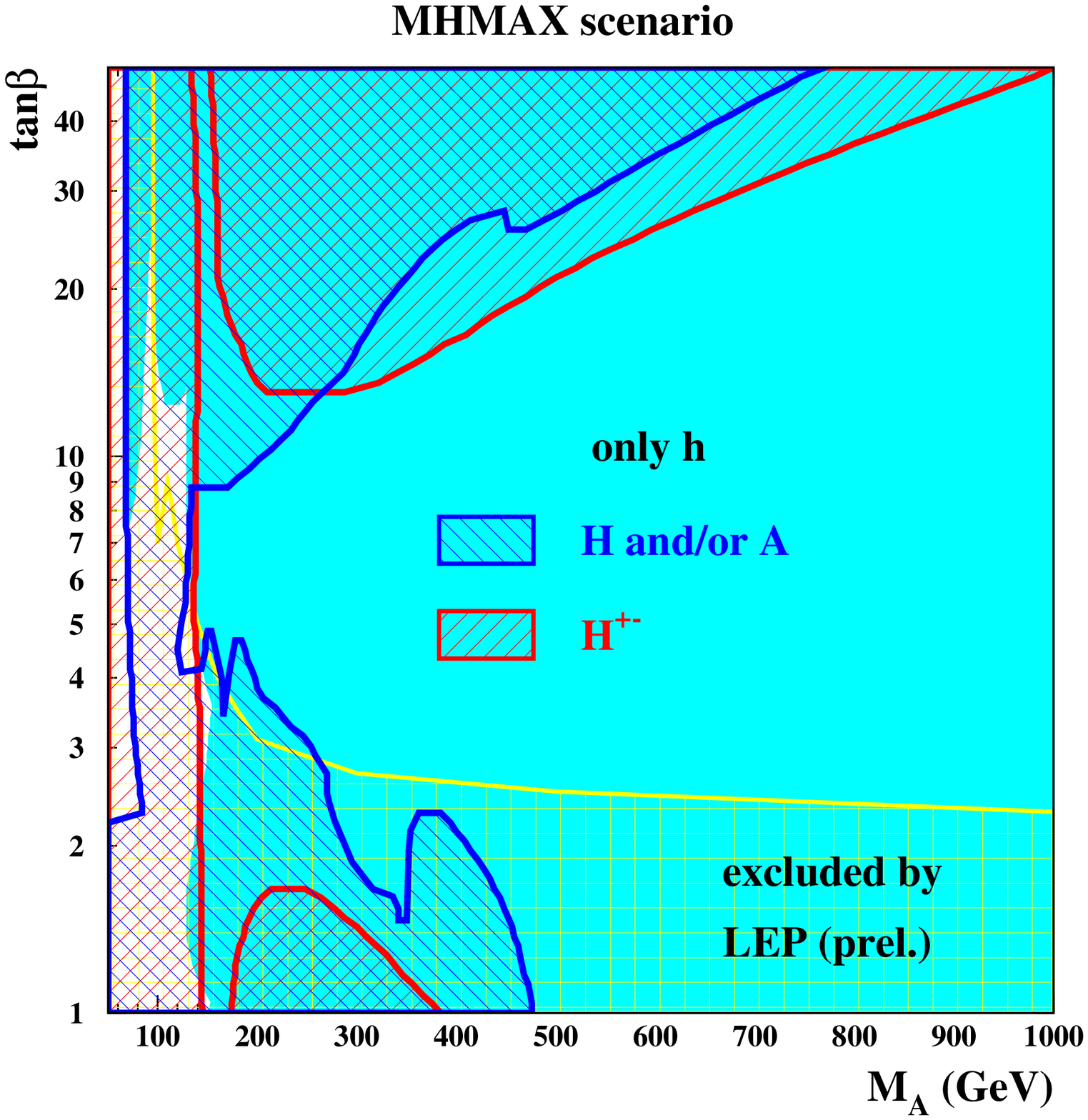,scale=.35}}
	  \put(310,60){ATLAS preliminary}
	\end{picture}
	\caption{ATLAS discovery potential for the light MSSM Higgs state $h$ with 30\,$\rm fb^{-1}$ (left); overall discovery potential for light and heavy neutral and charged MSSM Higgs bosons for 300\,$\rm fb^{-1}$ (right).}
	\label{fig:overall}
\end{figure}
The overall discovery potential of ATLAS is similar in all four benchmark scenarios. In the following, only some important aspects are discussed (often using the $M_h$-max scenario as an example), more details can be found in~\cite{SCH}. For an integrated luminosity of 30\,$\rm fb^{-1}$ the discovery potential for the {\bf light Higgs boson} $h$ is dominated by the VBF channel with $h\rightarrow \tau \tau$ in all scenarios considered. The results in the $M_h$-max scenario are shown in figure~\ref{fig:overall} (left). The VBF channel with $h\rightarrow \tau \tau$ covers most of the parameter space left over from LEP studies. In the small $\alpha$ scenario the effect of suppressed branching ratios into $\tau$ leptons is mainly important for $\tan\beta>20$ and $200\,{\rm GeV} < M_A < 300\,{\rm GeV}$. In this area the reduced discovery potential via the $\tau$ channels is nicely complemented by the $h$ decays to gauge bosons from VBF or GGF. 
For 300\,$\rm fb^{-1}$, also the channels $h\rightarrow \gamma \gamma$ and $h\rightarrow ZZ\rightarrow 4l$ and $tth$ with $h\rightarrow bb$ contribute significantly. For all benchmark scenarios discovery is possible via several channels in large parts of the parameter space allowing the determination of parameters of the Higgs sector. The discovery potential for the {\bf heavier neutral Higgs boson states} is given by the associated production with $b$ quarks and the decay into a pair of muons and tau leptons. These channels cover the regions of high $\tan\beta$. The production of charged Higgs bosons can be observed from top quark decays for $M_{H^{\pm}}<170$\,GeV and from gluon bottom fusion for $M_{H^{\pm}}>180$\,GeV. The overall discovery potential for Higgs boson states in the $M_h$-max scenario using 300\,$\rm fb^{-1}$ is summarized in figure~\ref{fig:overall} (right). In the whole parameter space at least one Higgs boson can be observed and for a significant part more than one Higgs boson can be discovered allowing to distinguish between the Higgs sector of the SM and its MSSM extension via direct observation. However a large area at intermediate $\tan\beta$ is left where only the light Higgs boson $h$ can be discovered. In this area the measurement of e.g. ratios of branching ratios in the same production mode of the $h$ may allow to distinguish the MSSM and SM Higgs sectors. Studies investigating this possibility are ongoing.

\section{Conclusion}
An updated evaluation of the MSSM Higgs discovery potential of ATLAS based on the most recent calculations for masses and branching ratios in four benchmark scenarios has been discussed. In the whole parameter space at least one Higgs boson can be discovered. The lightest Higgs state can often be discovered in multiple search channels, allowing maybe an indirect discrimination whether the SM or MSSM is realised in nature. The heavier Higgs state can only be discovered in parts of the parameter space.

\end{document}